\documentclass[12pt,oneside, a4paper]{article}
\ifx\pdfoutput\undefined
\usepackage[dvips,bookmarks=false]{hyperref}	% This is for arXiv.org
\else
\usepackage{hyperref}	% This is for pdftex
\fi
\hypersetup{colorlinks,bookmarksopen,bookmarksnumbered,citecolor=blue,
linkcolor=black,pdfstartview=FitH,urlcolor=blue}

\usepackage{graphicx}
\usepackage{amssymb}
\usepackage{cite}
\usepackage{bm}
\usepackage{indentfirst}
\usepackage{amsmath}
\usepackage{hhline}
\begin{document}

\begin{titlepage}

\begin{flushright}
KIAS-P14068\\
LPT-Orsay-14-86
\end{flushright}

\begin{center}

\vspace{1cm}
{\large\bf 
Effect of Degenerate Particles on\\ Internal Bremsstrahlung of Majorana
 Dark Matter
}
\vspace{1cm}

\renewcommand{\thefootnote}{\fnsymbol{footnote}}
Hiroshi Okada$^1$\footnote[1]{hokada@kias.re.kr} 
and 
Takashi Toma$^2$\footnote[2]{takashi.toma@th.u-psud.fr}
\vspace{5mm}

{\it%
$^1${School of Physics, KIAS, Seoul 130-722, Republic of Korea}\\
$^2${Laboratoire de Physique Th\'eorique, CNRS - UMR 8627, Universit\'e
 de Paris-Sud 11 F-91405 Orsay Cedex, France}
}

\vspace{8mm}

\abstract{
Gamma-rays induced by annihilation or decay of dark matter can be its
 smoking gun signature. In particular, gamma-rays generated by internal 
 bremsstrahlung of Majorana and real scalar dark matter is promising
 since it can be a leading emission of sharp gamma-rays. 
 However in the case of Majorana dark matter, 
 its cross section for internal bremsstrahlung cannot be large enough to
 be observed by future gamma-ray experiments 
 if the observed relic density is assumed to be thermally produced.
 In this paper, we introduce some degenerate particles with Majorana dark matter,
 and show they lead enhancement of the cross section.
 As a result, increase of about one order of magnitude for the cross
 section is possible without conflict with the observed relic density,
 and it would be tested by the future gamma-ray experiments
 such as GAMMA-400 and Cherenkov Telescope Array (CTA).
 In addition, the constraints of perturbativity, positron observation by 
 the AMS experiment and direct search for dark matter are discussed.
 }

\end{center}
\end{titlepage}

\renewcommand{\thefootnote}{\arabic{footnote}}
\setcounter{footnote}{0}

\setcounter{page}{2}

%%%%%%%%%%%%%%%%%%%%%%%%%%%%%%%%%%%%%%
\section{Introduction}
The existence of dark matter (DM) is crucial from cosmological observations,
however its mass and interaction are not known yet.
There are a lot of theoretical
DM candidates such as Weakly Interacting Massive Particle
(WIMP), axion, gravitino and asymmetric DM. In particular, WIMP
is the most promising DM candidate, and experiments are focusing on WIMP detection. 

One of the ways to identify nature of DM is indirect detection in
which experiments observe such as gamma-rays, 
neutrinos, positrons and anti-protons coming from the galaxy. 
Since DM may annihilate or decay into
the Standard Model (SM) particles, if observational
excess is found over background estimation, it can be a DM signature. 
Focusing on gamma-rays, known astrophysical sources of gamma-rays induce only
smooth spectra, while the DM annihilation or decay may produce a sharp gamma-ray
peak. Thus exploring such a line shape spectrum coming from the galaxy 
is extremely important, and DM models which can induce a strong
sharp gamma-ray are fascinating from experimental
viewpoint. 
In fact, the gamma-ray excesses from the galactic center have been
reported by several papers~\cite{Bringmann:2012vr, Weniger:2012tx}
and~\cite{Hooper:2011ti, Abazajian:2014fta, Carlson:2014nra}. 
The former implies the DM mass to be $130$-$135~\mathrm{GeV}$
with its partial annihilation cross section
$\sigma{v}\sim10^{-27}~\mathrm{cm^3/s}$ into gamma-rays. 
However this signal has disappeared after the dedicated
analysis by the Fermi Collaboration~\cite{Ackermann:2013uma}. 
The latter has found rather mild excess around $10~\mathrm{GeV}$ which
could be caused by $30$-$60~\mathrm{GeV}$ of the DM
mass with a specific annihilation mode such as a pair of bottoms or taus. 
The partial cross section should be
$\sigma{v}\sim10^{-26}~\mathrm{cm^3/s}$ that is very close to the
cross section for thermal production of the observed relic density. 

Representative processes inducing sharp gamma-rays are two-body
annihilations such as $\chi\chi\to\gamma\gamma$ and
$\chi\chi\to\gamma{Z}$.
Since the energy of the photon is kinematically determined as
$E_\gamma=m_\chi$ for $\gamma\gamma$ channel and
$E_\gamma=m_\chi\left(1-m_Z^2/(4m_\chi^2)\right)$ for $\gamma{Z}$
channel, gamma-rays become monochromatic. 
However, the cross sections for these processes are typically expected
to be small 
because of loop suppression. Another process of sharp gamma-rays is
internal bremsstrahlung
$\chi\chi\to{f}\overline{f}\gamma$~\cite{Bergstrom:1989jr,
Flores:1989ru, Bringmann:2007nk,
Ciafaloni:2011sa, Barger:2011jg, Garny:2013ama,
Kopp:2014tsa}.\footnote{There is also a model which produces a 
characteristic box-type gamma-ray spectrum mediated by a light
axion-like scalar~\cite{Ibarra:2012dw, Ibarra:2013eda}.} 
When the two-body process $\chi\chi\to{f}\overline{f}$ is
chirally-suppressed, the internal bremsstrahlung process becomes important for
the gamma-ray source of DM. 
The cross section for this process becomes larger than monochromatic
gamma-ray processes since this is three-body process at tree level. 
The gamma-ray spectrum of internal bremsstrahlung can be very 
sharp if mass of the mediate particle is not far from the
DM mass. 
%\footnote{Astrophysical
%uncertainties may make it difficult to see the sharp
%spectral feature of internal bremsstrahlung.}  
This is because fermion and anti-fermion with soft energy is produced in the final
state and the most energy of the initial state is taken away by the other two
particles. As a result, the energy of photon is almost fixed to half
of the total energy of the initial state
$2m_\chi$.
In particular for real scalar DM, stronger sharp gamma-rays can be
induced than the Majorana DM case due to d-wave suppression of the
two-body process $\chi\chi\to f\overline{f}$~\cite{Toma:2013bka,
Giacchino:2013bta, Ibarra:2014qma, Giacchino:2014moa}. 
In the case of Majorana DM, it is known that the
cross section for internal bremsstrahlung $\chi\chi\to
f\overline{f}\gamma$ cannot be large enough to be detected in near
future if thermal production of DM is concerned~\cite{Garny:2013ama}. 
This is because suppression for $\chi\chi\to f\overline{f}$ is p-wave and 
once the interaction strength is fixed in order to
obtain the observed relic density of DM, the cross section for internal
bremsstrahlung is also determined. 
It has been claimed that the gamma-ray spectrum of internal
bremsstrahlung can reproduce nicely the $130$-$135~\mathrm{GeV}$ line 
mentioned above, however this is not compatible with thermal relic
density of DM~\cite{Bringmann:2012vr}. 
One should note that the internal bremsstrahlung process is also constrained by
AMS positron observation since a pair of high energy electron and
positron is also produced~\cite{Aguilar:2013qda, Bergstrom:2013jra}. 

In this paper, we consider enhancement of the internal bremsstrahlung
process for Majorana DM due to increase of DM effective degrees of
freedom taking into account degenerate particles with DM. 
Assuming that the degenerate particles weakly interact with
the SM particles, the effective annihilation cross section for relic
density would be smaller than the DM self-annihilation cross section. 
Thus the interaction of DM should be
larger in order to satisfy the observed DM relic density. As a
result, it is expected that the cross section for internal
bremsstrahlung is also increased. 
Moreover we will consider the constraints from perturbativity, positron
observation by the AMS experiment and direct
detection experiments. 
A similar work has been done for a Higgsino- or
Wino-like neutralino DM in a supersymmetric
model~\cite{Profumo:2006bx}. In this reference, the authors have considered
co-annihilations with sleptons. Because of increase 
of DM couplings for the correct relic density due to weak
co-annihilations, annihilation rates into neutrinos and anti-protons 
have been enhanced as indirect detection signals of DM. Our work in this
paper is an application to internal bremsstrahlung in a simplified
Majorana DM model.

This paper is organized as follows. 
In the next section, we quantitatively review the standard
internal bremsstrahlung of a single Majorana DM, and 
explicitly show the cross section for the process cannot be large enough to be
detected by future gamma-ray experiments. 
In Section~\ref{sec:3}, we introduce several degenerate particles with
DM and show enhancement of the cross section is derived due to the
degenerate particles. Our conclusions are given in Section~\ref{sec:4}.

\section{Standard Internal bremsstrahlung}
\label{sec:2}
\subsection{Interactions and cross sections}

We consider the leptonic Yukawa interaction 
\begin{equation}
\mathcal{L}=y\varphi\overline{\chi}P_R\ell+\mathrm{H.c.},
\end{equation}
where the Yukawa coupling $y$ is assumed to be real for simplicity, $\chi$ is a singlet Majorana
DM in the SM gauge group, $\varphi$ is an electromagnetically 
charged scalar mediator with hypercharge $Y_\varphi=1$ and $\ell$ is a
right-handed charged lepton. 
Although DM can couple to three generations of the charged leptons in
general, we consider an interaction with only one generation for simplicity. 
DM is regarded to be stabilized by a $\mathbb{Z}_2$ symmetry. 
In our case, $\mathbb{Z}_2$ assignment should be odd for DM
$\chi$ and the mediator $\varphi$ and even for the charged lepton $\ell$. 

\begin{figure}[t]
\begin{center}
\includegraphics[scale=0.6]{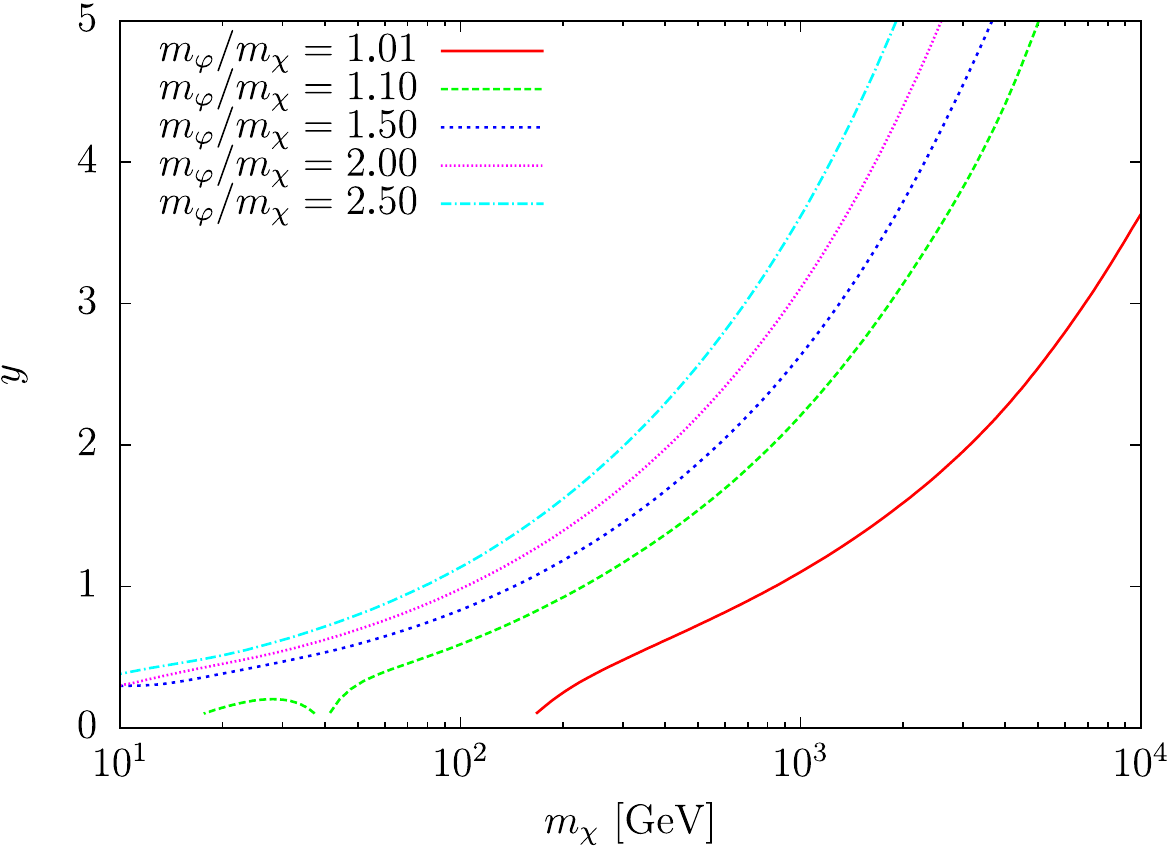}
\hspace{0.5cm}
\includegraphics[scale=0.6]{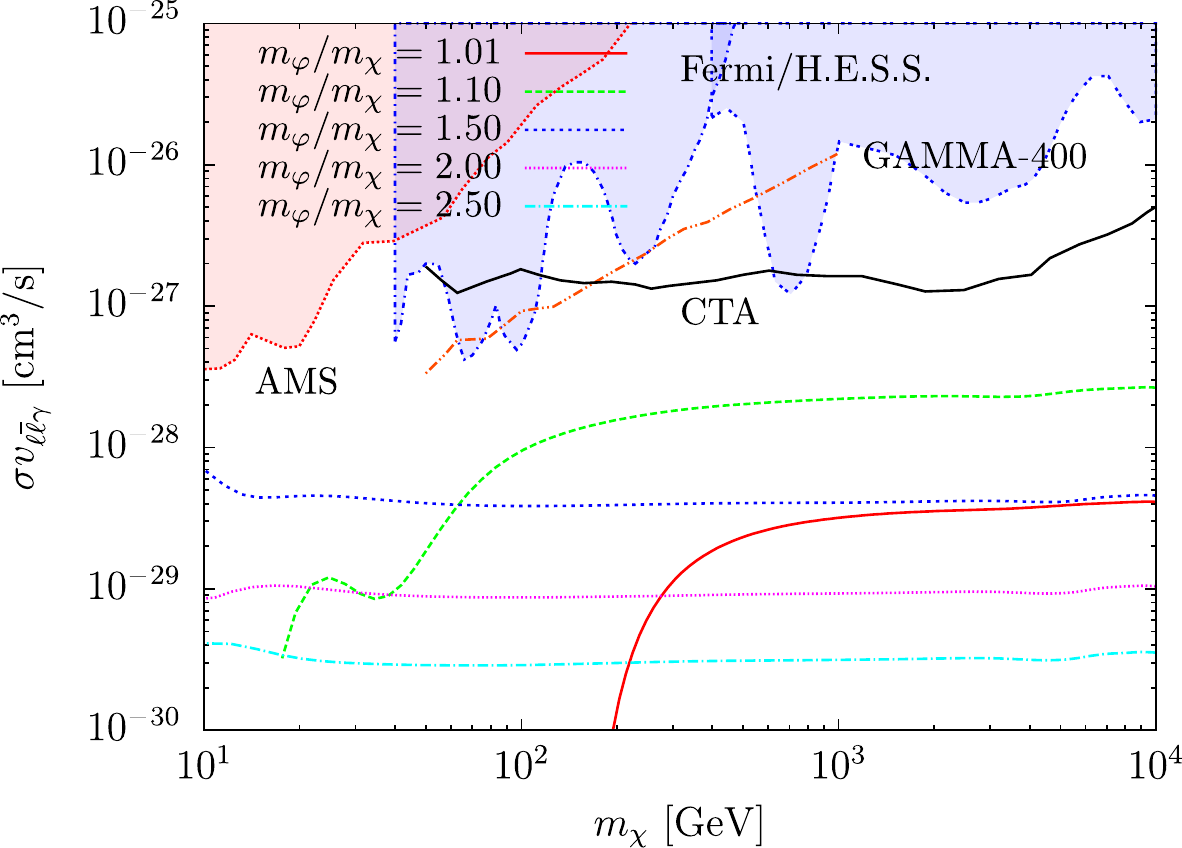}
\caption{Contours satisfying the measured DM relic density on
 ($m_\chi$, $y$) plane (left panel). Comparison
 of the cross section for internal bremsstrahlung with gamma-ray
 experiments and the AMS positron observation (right panel).} 
\label{fig:yukawa}
\end{center}
\end{figure}

The thermal relic density of DM is determined by the annihilation cross
section into a pair of leptons through the Yukawa coupling, and should
satisfy the observed one $\Omega h^2\approx0.120$~\cite{Ade:2013zuv}. 
When the DM mass is much heavier than the charged lepton mass $m_\ell/m_\chi\ll1$, the
annihilation cross section into $\ell\overline{\ell}$ is calculated as  
\begin{equation}
\sigma{v}_{\ell\overline{\ell}}=\frac{y^4}{48\pi m_\chi^2}
\frac{1+\mu^2}{\left(1+\mu\right)^4}v^2,
\label{eq:p-wave}
\end{equation}
where $\mu=m_\varphi^2/m_\chi^2$ is the mass ratio between DM
and the mediator, and $v$ is the relative velocity of DM. 
There is no s-wave due to chiral suppression. 
The required strength of the Yukawa coupling for the several fixed mass
ratios $m_\varphi/m_\chi$ is shown in the left panel in
Fig.~\ref{fig:yukawa}. As one can see from the figure, $\mathcal{O}(1)$ Yukawa
coupling is needed for the measured relic density. 
When the mass ratio is
$m_\varphi/m_\chi=1.01$, the required Yukawa coupling becomes rather
small since the co-annihilation with the mediator is strong. 
In this case, the DM mass being less than $170~\mathrm{GeV}$ gives too small
relic density and cannot satisfy the measured relic density.
There is a small dip around $m_\chi\approx40~\mathrm{GeV}$ for
$m_\varphi/m_\chi=1.1$. This is because DM and the mediator $\varphi$
are $10\%$ degenerate, and the effective annihilation cross
section of DM becomes large due to the $\varphi$ self-annihilation channel: 
$\varphi^\dag\varphi\to f\overline{f}$ mediated by the $s$-channel $Z$ boson
where $f$ is a SM fermion.  
The region of the DM mass more than a few TeV tends to be ruled
out by perturbativity of the Yukawa coupling. 

As one can see from Eq.~(\ref{eq:p-wave}), the cross section is
suppressed by the relative velocity squared (p-wave) due to 
the helicity suppression. Because of that, this channel is extremely
suppressed in current times since the averaged DM relative
velocity is roughly $v\sim10^{-3}$. 
As a result, three-body process $\chi\chi\to \ell\overline{\ell}\gamma$
becomes important for indirect detection of DM~\cite{Bergstrom:1989jr,
Flores:1989ru, Bringmann:2007nk,
Ciafaloni:2011sa, Barger:2011jg, Garny:2013ama,
Kopp:2014tsa}. 
There are three diagrams contributing to the process. 
The total amplitude is separated to Final State
Radiation (FSR) and Virtual Internal Bremsstrahlung (VIB) parts in a
gauge invariant way~\cite{Bringmann:2007nk, Ciafaloni:2011sa}.
Among them, the differential cross section of the FSR contribution
is model-independently given by 
\begin{equation}
\frac{d\sigma{v}_{\ell\overline{\ell}\gamma}^{\mathrm{FSR}}}{dx}=
\sigma{v}_{\ell\overline{\ell}}\frac{\alpha_{\mathrm{em}}}{\pi}
\frac{(1-x)^2+1}{x}\log\left(\frac{4m_\chi^2\left(1-x\right)}{m_\ell^2}\right),
\end{equation}
where $x=E_\gamma/m_\chi$.
Since this cross section is proportional to the chirally-suppressed two-body cross section
$\sigma{v}_{\ell\overline{\ell}}$, the FSR contribution to the gamma-ray
source is negligible. 
Therefore the other VIB contribution becomes dominant.\footnote{The interference term with the
amplitude for FSR is also negligible in the region of $E_\gamma\sim m_\chi$.} 
This process may induce sharp gamma-ray spectrum when the $\varphi$ mass is not
far from the DM mass ($\mu\lesssim2$). 
In the limit of $m_\ell/m_\chi\to0$, the cross section for internal
bremsstrahlung $\chi\chi\to \ell\overline{\ell}\gamma$ is calculated
as~\cite{Bringmann:2007nk} 
\begin{eqnarray}
\sigma{v}_{\ell\overline{\ell}\gamma}\hspace{-0.2cm}&=&\hspace{-0.2cm}
\frac{y^4\alpha_{\mathrm{em}}}{64\pi^2m_\chi^2}
\left[
\left(\mu+1\right)\left\{
\frac{\pi^2}{6}-\log^2\left(\frac{\mu+1}{2\mu}\right)
-2\mathrm{Li}_2\left(\frac{\mu+1}{2\mu}\right)
\right\}
+\frac{4\mu+3}{\mu+1}\right.\nonumber\\
&&\hspace{1.5cm}\left.
+\frac{(4\mu+1)(\mu-1)}{2\mu}\log\left(\frac{\mu-1}{\mu+1}\right)
\right].
\label{eq:ib}
\end{eqnarray}

There are also the loop-induced channels $\chi\chi\to\gamma\gamma$ and $\chi\chi\to\gamma
Z$ as the other gamma-ray sources. In this model, the annihilation
cross section for $\chi\chi\to\gamma\gamma$ is calculated as~\cite{Giacchino:2014moa}
\begin{equation}
\sigma{v}_{\gamma\gamma}=\frac{y^4\alpha_\mathrm{em}^2}{256\pi^3
 m_\chi^2}
\left|\mathrm{Li}_2\left(\frac{1}{\mu}\right)
-\mathrm{Li}_2\left(-\frac{1}{\mu}\right)\right|^2.
\end{equation}
This cross section is typically smaller than the cross section for
internal bremsstrahlung at $\mu\sim1$ due to the loop suppression as have discussed in
ref.~\cite{Giacchino:2014moa}, and the cross section for
$\chi\chi\to\gamma{Z}$ is also small.
If DM $\chi$ has the other interactions with
the SM particles and annihilates as $\chi\chi\to b\overline{b}$, $W^+W^-$, $ZZ$,
continuous broad gamma-rays are produced and it should be included in
total gamma-ray sources from DM. 
In particular, if the mediator $\varphi$ has a $SU(3)_c$
charge, anti-protons are produced due to gluon internal bremsstrahlung
$\chi\chi\to q\overline{q}g$ through the interaction
$y\varphi\overline{\chi}q$ which is obtained by replacing the charged
lepton $\ell$ with a quark $q$~\cite{Asano:2011ik}. 
This gluon internal bremsstrahlung would give a stronger constraint than
photon internal bremsstrahlung. 
However in this simplified model, such annihilations do not occur
since the $SU(3)_c$ singlet mediator $\varphi$ is concerned.

The cross section for internal bremsstrahlung is numerically calculated
by using the
Yukawa coupling which satisfies the observed relic density as shown in the
right panel in Fig.~\ref{fig:yukawa}. 
In the region of $m_\chi\lesssim300~\mathrm{GeV}$ for
$m_\varphi/m_\chi=1.01$, since the co-annihilation and self-annihilation
of $\varphi$ are dominant, the Yukawa coupling needed for the observed relic 
density becomes small. Therefore, the
cross section for internal bremsstrahlung is also small. 
For $m_\varphi/m_\chi\approx1.10$, the cross section can be maximal, 
and for larger mass ratio, although the required Yukawa
coupling for the observed relic density can be sizable, 
the cross section for internal bremsstrahlung remains small
since it decreases rapidly as
$\sigma{v}_{\ell\overline{\ell}\gamma}\propto \mu^{-4}$ for
$\mu\gg1$.

\subsection{Gamma-ray bounds, prospects and AMS positron observation}
In order to set an upper bound for the cross section generating sharp
gamma-rays, the cross section is translated to gamma-ray flux and
specific target region of interest for observation is taken as followed
by ref.~\cite{Garny:2013ama}. 
The Fermi/H.E.S.S. limit has been obtained by assuming
an Einasto DM profile with appropriate parameter setting in
ref.~\cite{Garny:2013ama}. The Fermi and H.E.S.S. data were taken from search region $3$ Pass
$7$ SOURCE sample in ref~\cite{Weniger:2012tx} and the central Galactic
halo (CGH) region in ref.~\cite{Abramowski:2013ax} respectively.
As shown in the right panel of Fig.~\ref{fig:yukawa}, the blue
region is excluded by Fermi-LAT and H.E.S.S. at $95\%$
Confidence Level (CL)~\cite{Garny:2013ama}, and the prospected
sensitivities of the future gamma-ray experiments GAMMA-400 and
Cherenkov Telescope Array (CTA) are 
described by the brown and black
lines~\cite{Galper:2012fp, Bernlohr:2012we}. 
As one can see, the maximal cross section for sharp gamma-rays in the
case of standard internal bremsstrahlung is about
$\sigma{v}_{\ell\overline{\ell}\gamma}\approx2\times10^{-28}~\mathrm{cm^3/s}$ 
which is about one order of magnitude below the GAMMA-400 and CTA prospects. 

Since high energy charged leptons are also produced by internal
bremsstrahlung, the cross section for $\chi\chi\to\ell\overline{\ell}\gamma$ is
constrained by the AMS positron observation~\cite{Aguilar:2013qda}. 
The upper bound of the cross section for internal bremsstrahlung
$\chi\chi\to e^+e^-\gamma$ has been studied by in
ref.~\cite{Bergstrom:2013jra} where an Einasto DM profile is assumed. 
This upper bound is shown as the red region in the right panel of Fig.~\ref{fig:yukawa}.
Note that even if $\ell=\mu,\tau$, high energy electrons and positrons
are emitted from the $\mu,\tau$ decays. 
However since their energy is softer than that of direct produced $e^+e^-$, the
constraint for $e^+e^-\gamma$ would be strongest and regarded as
conservative upper bound for $\chi\chi\to\ell\overline{\ell}\gamma$.

As we have discussed here, 
the cross section for the standard internal bremsstrahlung of Majorana
dark matter cannot reach the prospected sensitivity of the future
gamma-ray experiments. 
Hence we move on to our scenario to obtain a larger cross section for
internal bremsstrahlung in the next section.

\section{Internal Bremsstrahlung with Degenerate Particles}
\label{sec:3}
\subsection{Interactions and cross sections}

Now we consider $k$ degenerate Majorana fermions with DM. The Yukawa 
interaction is extended by 
\begin{equation}
{\cal L}= \sum_{i=1}^ky_{i} \varphi\overline{\chi_i}P_R\ell+{\rm H.c.},
\end{equation}
where $\chi_1$ is the lightest and regarded as DM with the mass
$m_1\equiv m_\chi$,
$\chi_i~(i\neq1)$ is assumed to be degenerate with DM whose
mass is denoted as $m_i~(i\neq1)$. 
In general, we can choose the diagonal base for the mass matrix of $\chi_i$. 
When DM is degenerate with these particles,
co-annihilation effect has to be taken into
account to evaluate thermal relic density of DM. 
Following ref.~\cite{Griest:1990kh}, the effective annihilation
cross section is given by 
\begin{equation}
\sigma_{\mathrm{eff}}v=\sum_{i=1}^k\sum_{j=1}^k\frac{g_{i}g_{j}}
{g_{\mathrm{eff}}^2}\sigma_{ij}v
\left(1+\Delta_i\right)^{3/2}\left(1+\Delta_j\right)^{3/2}
e^{-(\Delta_i+\Delta_j)x},
\label{eq:coann}
\end{equation}
where $g_{\mathrm{eff}}$ is the effective degrees of freedom 
\begin{equation}
g_{\mathrm{eff}}=\sum_{i=1}^kg_{i}\left(1+\Delta_i\right)^{3/2}e^{-\Delta_ix},
\end{equation}
and $\Delta_i\equiv (m_i-m_1)/m_1$ is the mass difference between DM
and the other degenerate particles, $g_i=2$ is the degrees of freedom
for each Majorana particle $\chi_i$, $x=m_1/T$ and $\sigma_{ij}v$ is
(co-)annihilation cross section between $i$ and $j$. If a lot of degenerate 
particles exist, the effective degrees of freedom $g_\mathrm{eff}$
increases. In addition, the Yukawa coupling
$y_i~(i\neq1)$ is larger than $y_1$, its effective annihilation cross section
$\sigma_\mathrm{eff}{v}$ also becomes large and the relic density of the
DM is much reduced. 
On the other hand, if the Yukawa coupling $y_i~(i\neq1)$ is smaller than
$y_1$, the
effective annihilation cross section becomes smaller than the standard
cross section of DM for non-degenerate case ($k=1$). Thus to compensate the lack
of the cross section, a larger coupling of DM $y_1$ is
required to satisfy the observed relic density. 
We consider such a small Yukawa coupling $y_i~(i\neq1)$ for the degenerate fermions
to increase the strength of the DM Yukawa coupling $y_1$. 
The cross section for internal bremsstrahlung is given by the same
formula with Eq.~(\ref{eq:ib}) and replacing $y\to y_1$.
%%%%%%%%%%%%%%%%%%%%%%%%%%%%%%%%%%%
\begin{figure}[tbc]
\begin{center}
\includegraphics[scale=0.6]{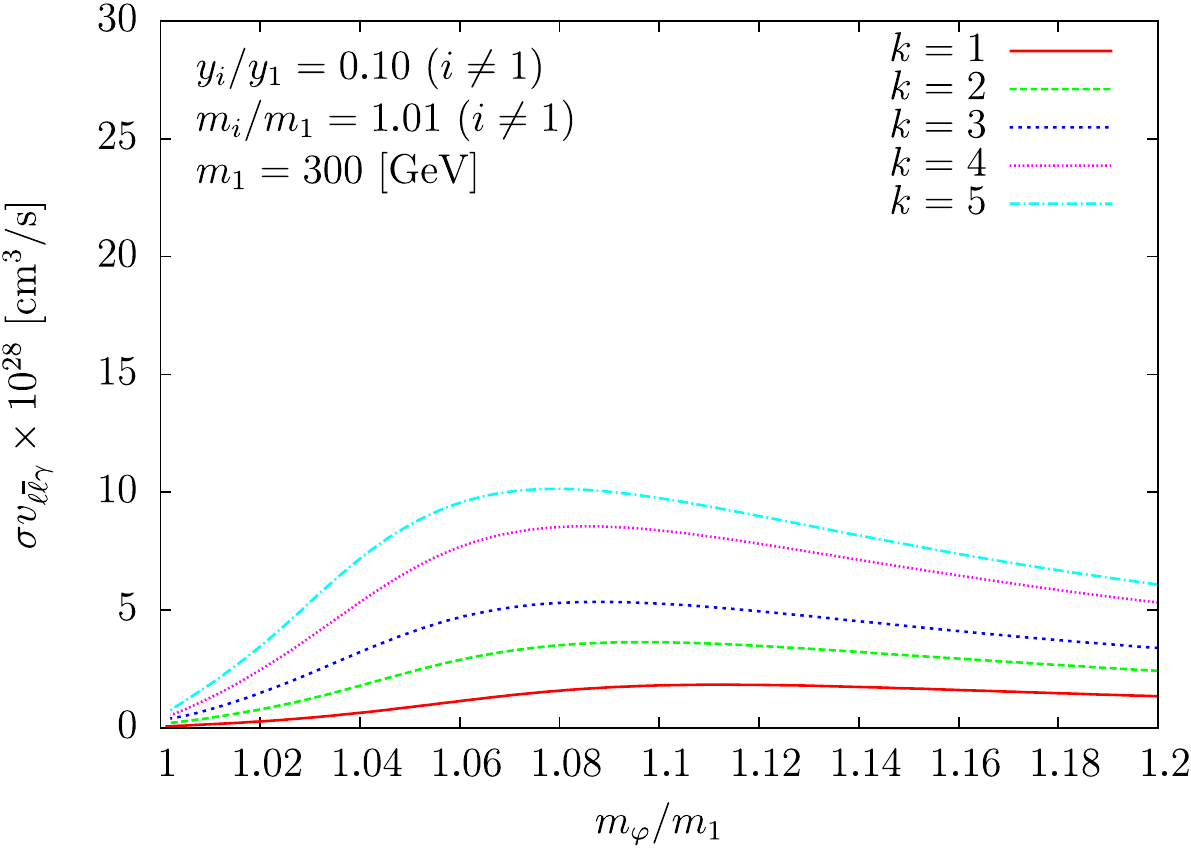}
\hspace{0.5cm}
\includegraphics[scale=0.6]{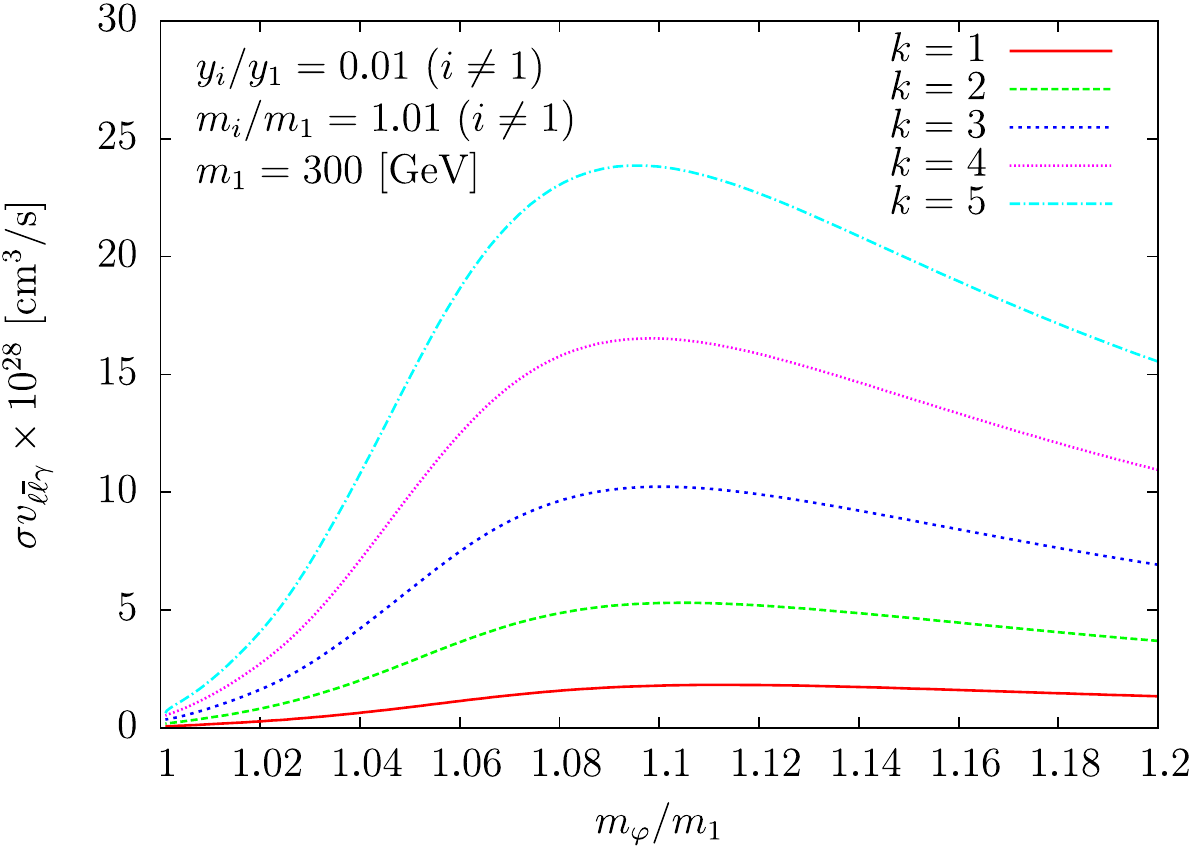}
\caption{$m_\varphi/m_1$ dependence of the cross section for internal
 bremsstrahlung. 
 The strength of the Yukawa coupling $y_i~(i\neq1)$ compared to $y_1$ is
 taken as $y_i/y_1=0.10$ (left panel) and $y_i/y_1=0.01$ (right panel).} 
\label{fig:coanni}
\end{center}
\end{figure}
%%%%%%%%%%%%%%%%%%%%%%%%%%%%%%%%%%%

%%%%%%%%%%%%%%%%%%%%%%%%%%%%%%%%%%%
\begin{figure}[t]
\begin{center}
\includegraphics[scale=0.6]{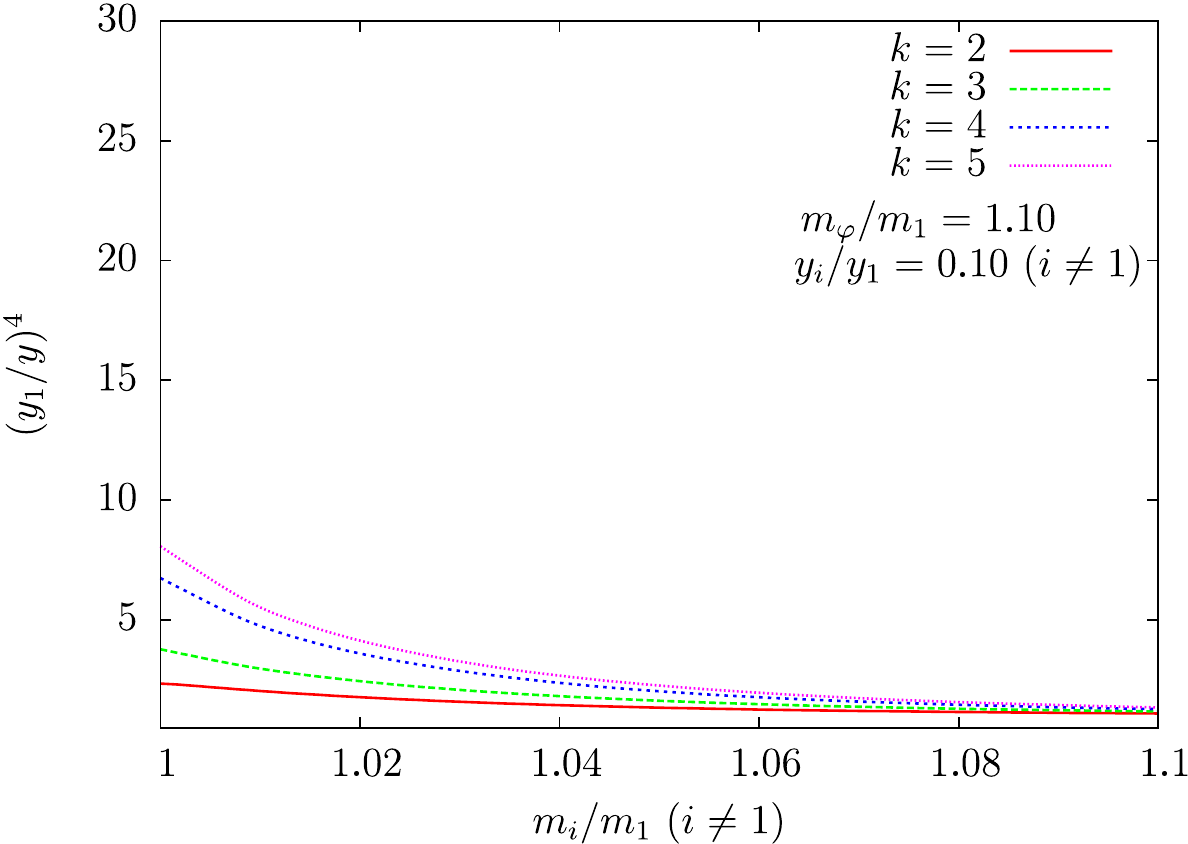}
\hspace{0.2cm}
\includegraphics[scale=0.6]{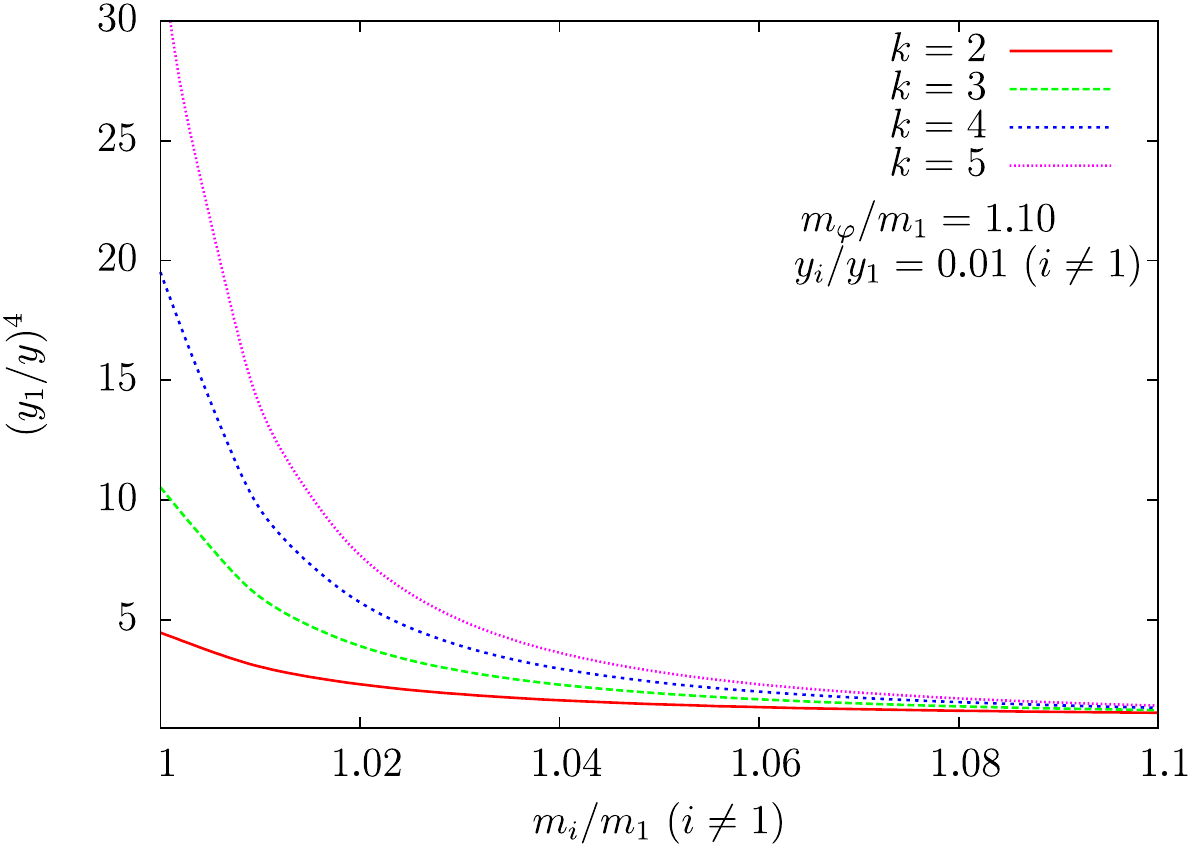}\\
\includegraphics[scale=0.6]{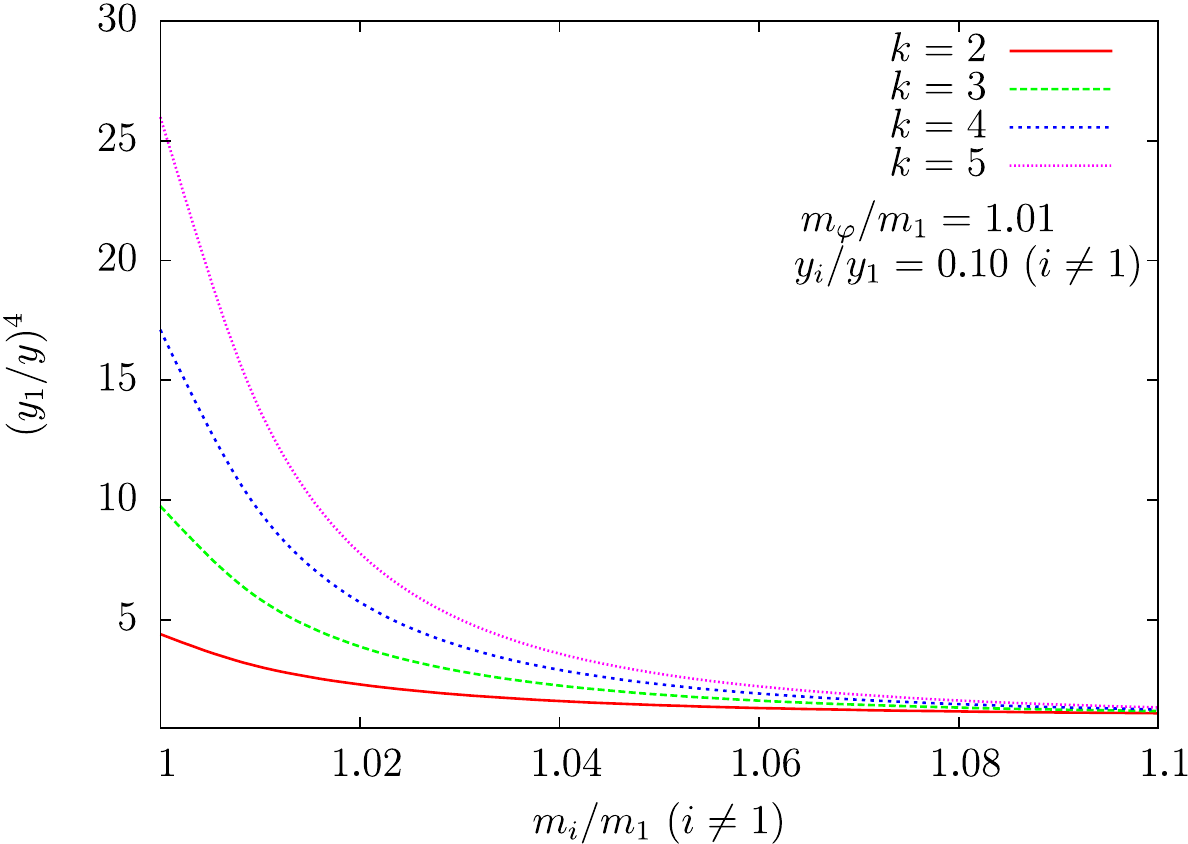}
\hspace{0.2cm}
\includegraphics[scale=0.6]{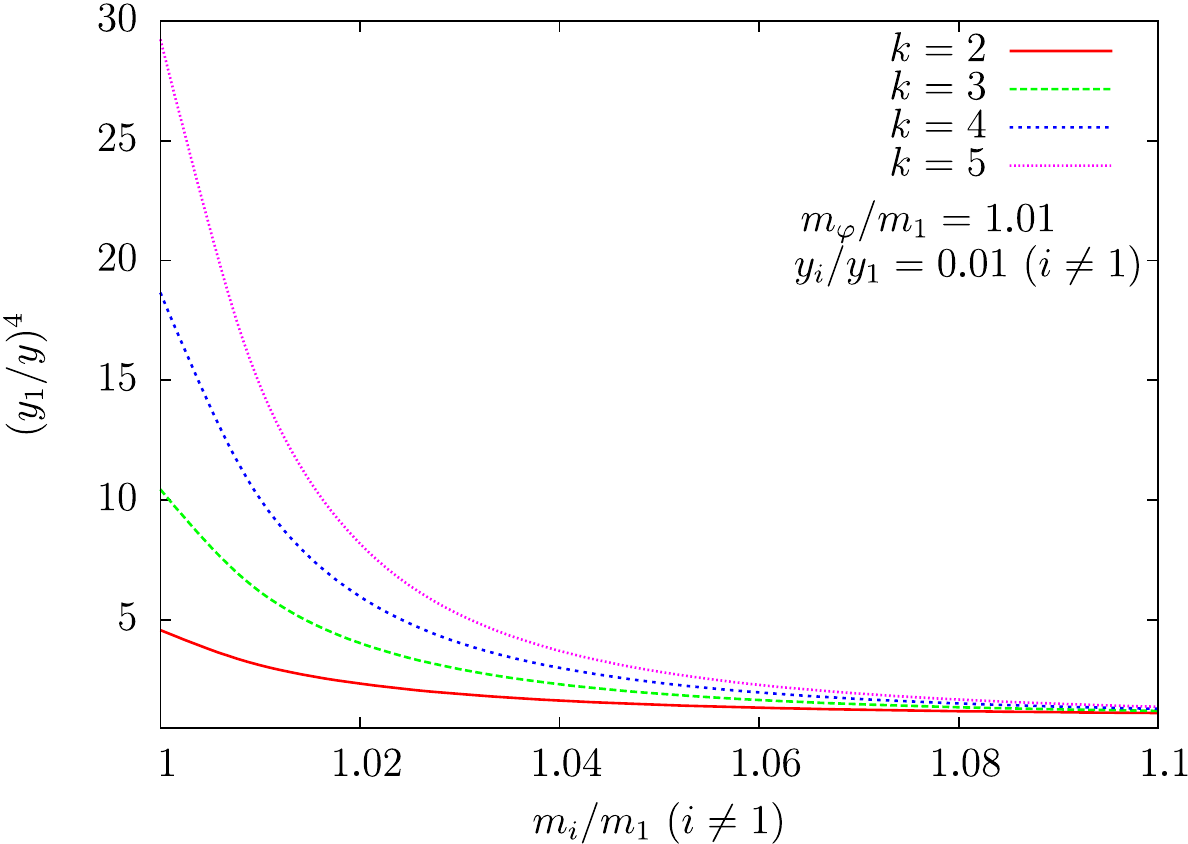}
\caption{Mass degeneracy dependence of the Yukawa coupling required for
 thermal relic density where $y$ in the y-axis is the Yukawa coupling
 for non-degenerate case, and all the degenerate fermion masses
 $m_i~(i\neq1)$ are assumed to be same.}
\label{fig:ib}\end{center}\end{figure}
%%%%%%%%%%%%%%%%%%%%%%%%%%%%%%%%%%%

We give some numerical results for effect of the DM Yukawa
coupling due to the degenerate Majorana fermions. 
The effective cross section has been calculated by
micrOMEGAs~\cite{Belanger:2013oya}. 
%%%
The cross section for internal bremsstrahlung obtained in the degenerate
scenario in
terms of the mass ratio between DM and the mediator ($m_\varphi/m_1$)
is shown in Fig.~\ref{fig:coanni}, where the DM mass $m_1$ is fixed to
be $300~\mathrm{GeV}$ and the mass ratio between the degenerate
particles and DM $m_i/m_1$ 
is fixed to $1.01$ as an example. We assumed that all the degenerate fermions have 
same masses for simplicity. The strength of the 
Yukawa coupling $y_i/y_1~(i\neq1)$ is fixed  to be $0.10$ in the left panel, and $0.01$ in the right panel.
The figure shows that larger enhancement of the cross section can be
obtained in the case of a smaller 
 $y_i/y_1$ and a larger number of degenerate particles as we expected. 
One also finds that a mild peak exists at a value of
$m_\varphi/m_1$ depending on the parameters. 
The position of the peak is in the range of
$1.06\lesssim m_\varphi/m_1\lesssim 1.12$. 
%%%

The dependence on the mass degeneracy $m_i/m_1$ of the DM Yukawa coupling required for
thermal DM production is shown in
Fig.~\ref{fig:ib} where the mass ratio $m_\varphi/m_1$ and the strength of the 
Yukawa coupling $y_i$ are fixed as shown in the figure. 
The mass ratio $m_\varphi/m_1$ should be $\mu\lesssim2$ to get
a sharp gamma-ray peak of internal bremsstrahlung. 
As one can see from the figure, when the number of the particles increases and they are
strongly degenerate, the required strength of the DM Yukawa coupling for the
observed relic density is enhanced. According to the figure, it is
possible to obtain one order of magnitude increase of $(y_1/y)^4$. 
We should note that for $m_\varphi/m_1=1.01$, large enhancement of the
Yukawa coupling occurs, however the cross section for internal
bremsstrahlung cannot be so large since the absolute value of the Yukawa
coupling is small. 
Although the DM mass must be fixed to
calculate thermal relic density, there is almost no dependence on the DM
mass. The reason is that we take the ratio of the cross sections
Eq.~(\ref{eq:p-wave}) and $\sigma_{11}{v}$ in Eq.~(\ref{eq:coann}) to
derive the ratio of the Yukawa coupling $(y_1/y)^4$, and the ratio of the
cross sections is dimensionless.

\subsection{Gamma-rays and the other constraints}

\begin{figure}[t]
\begin{center}
\includegraphics[scale=0.6]{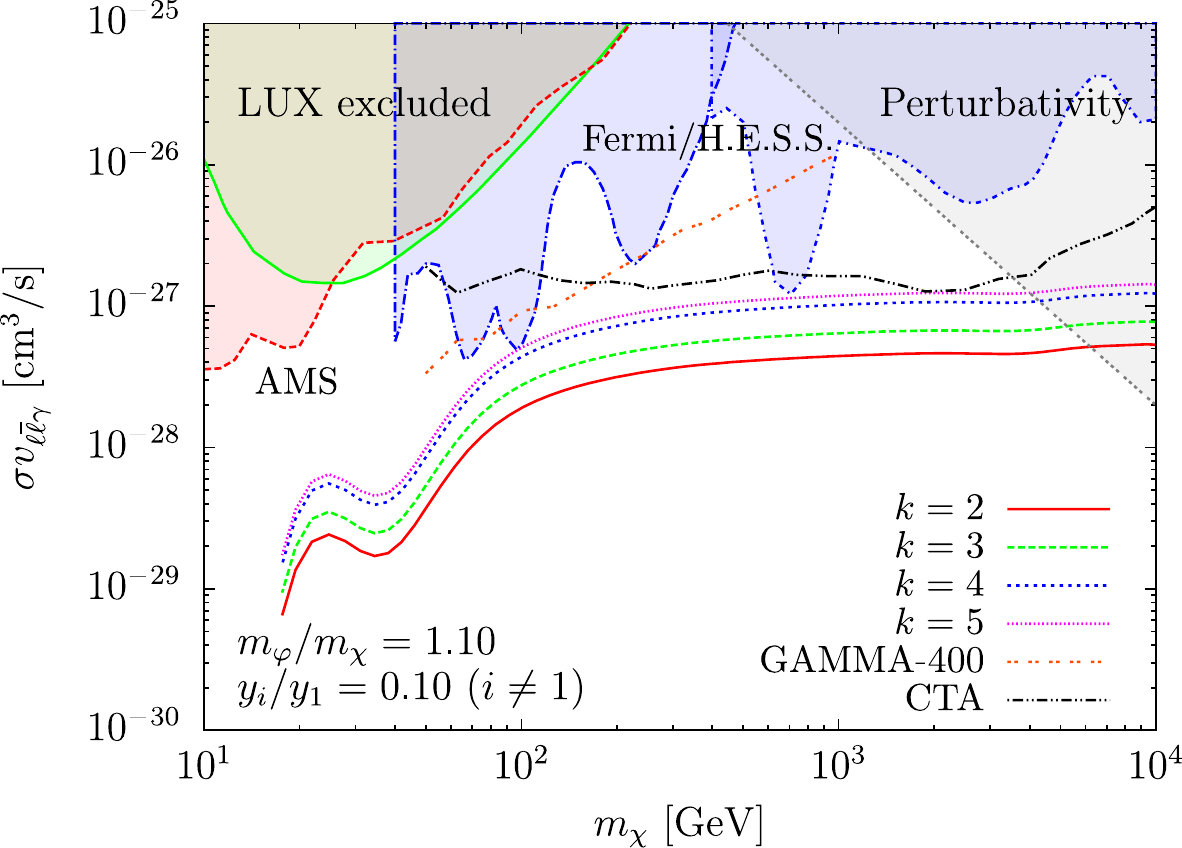}
\hspace{0.2cm}
\includegraphics[scale=0.6]{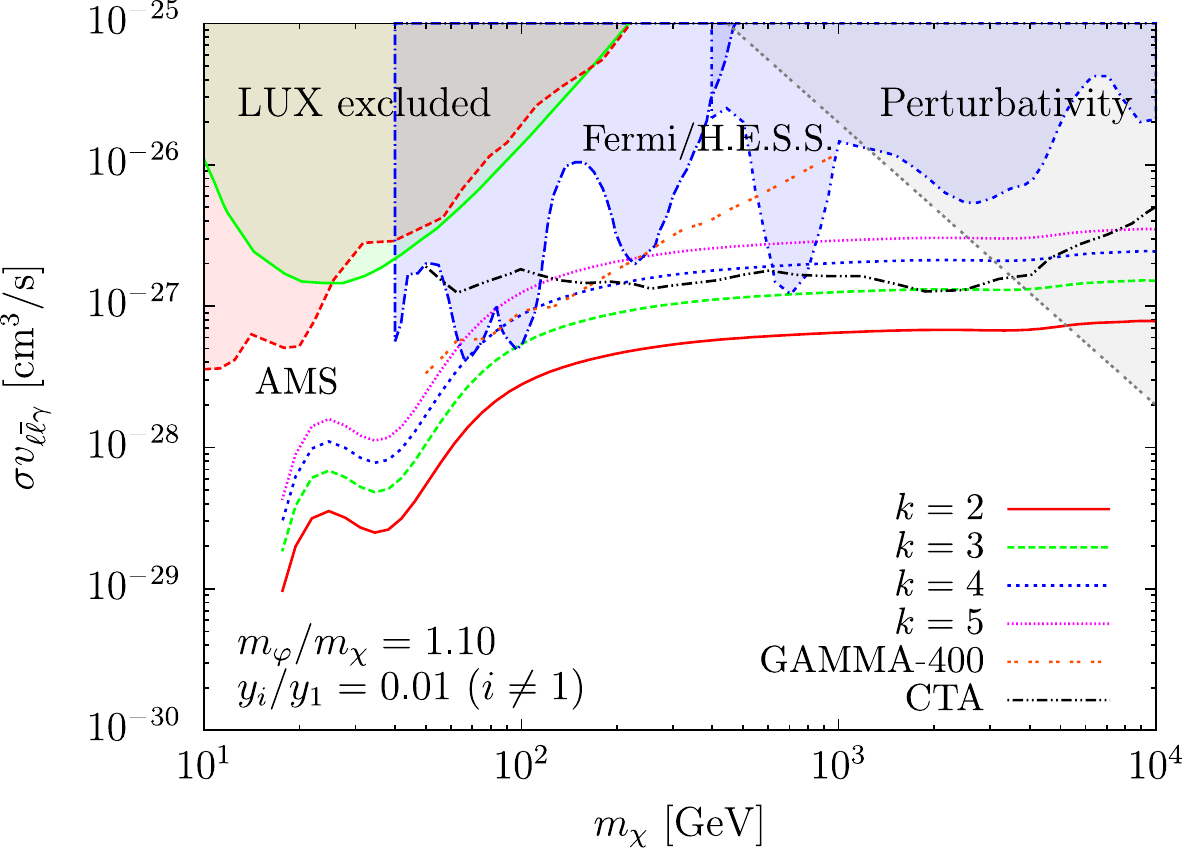}\\
\includegraphics[scale=0.6]{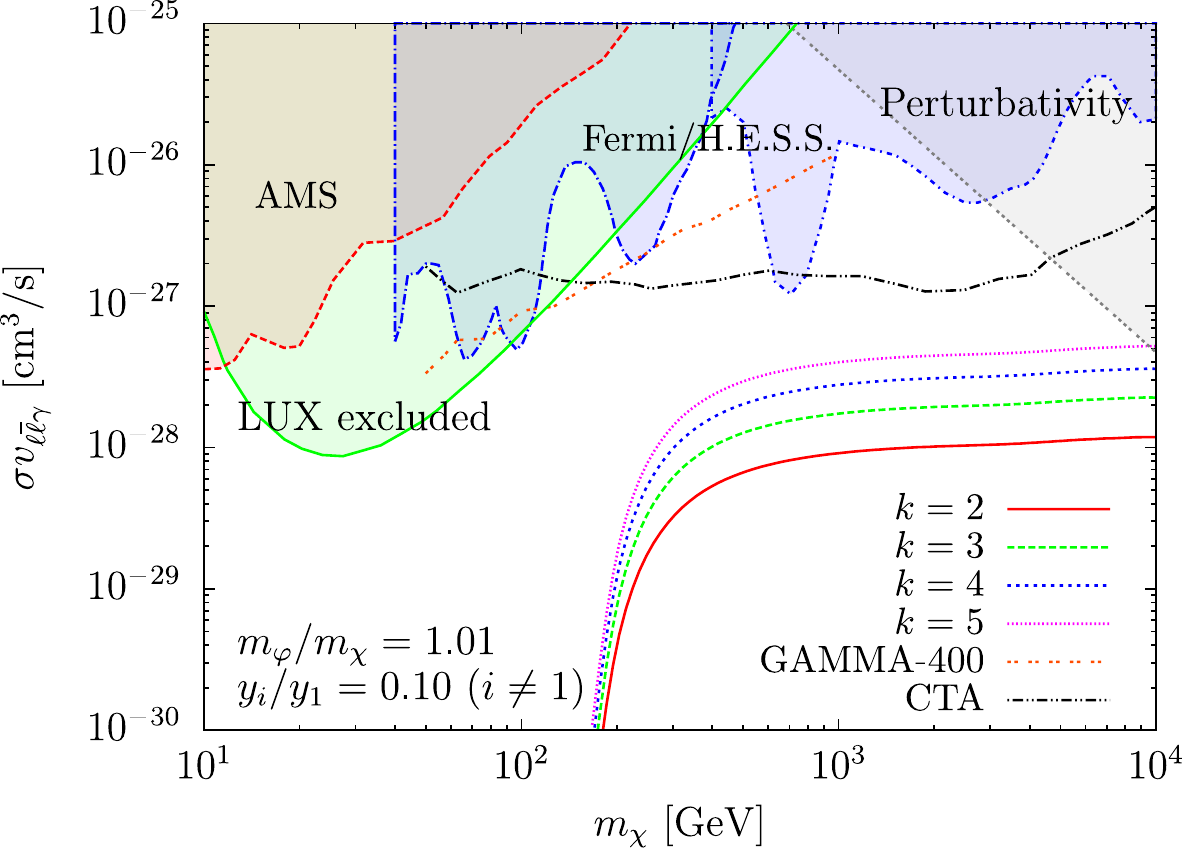}
\hspace{0.2cm}
\includegraphics[scale=0.6]{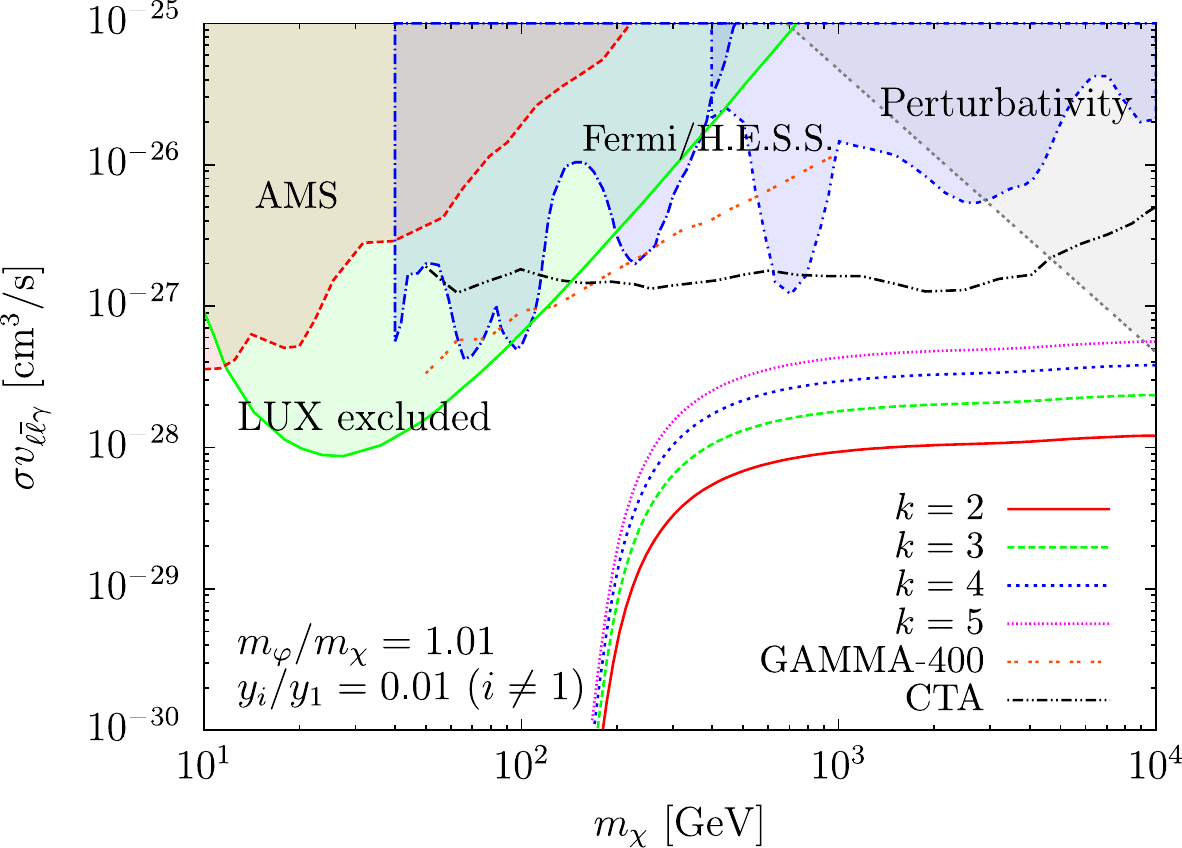}
\caption{Comparison of the cross section for internal bremsstrahlung with
 gamma-ray experiments where $m_\varphi/m_1$ and $y_i/y_1$ are fixed as
 in each figure, and $m_i/m_1=1.01$. The grey, green and red
 regions are excluded by perturbativity of the 
 Yukawa coupling $y_1$, the LUX experiment and the AMS positron
 observation respectively.}
\label{fig:exp}
\end{center}
\end{figure}

The comparison with the Fermi/H.E.S.S. bound and the future gamma-ray
experiments GAMMA-400 and CTA is shown in Fig.~\ref{fig:exp} 
where the mass ratio $m_\varphi/m_1$ and the Yukawa coupling of the
degenerate fermions are fixed as same as Fig.~\ref{fig:ib}. 
In particular, one can see that for $m_\varphi/m_1=1.10$ and
$y_i/y_1=0.01$ (upper right panel), the cross section for internal bremsstrahlung is 
largely enhanced and testable by GAMMA-400 and CTA~\cite{Galper:2012fp, Bernlohr:2012we}. For
$m_\varphi/m_1=1.01$, the cross section is rather smaller than the case for
$m_\varphi/m_1=1.10$ due to the strong co-annihilation with
$\varphi$.

Some other constraints are also shown in Fig.~\ref{fig:exp}.
The conservative perturbativity bound $y_1\leq\sqrt{4\pi}$ is taken into
account and the excluded region is shown in
the figure as grey color. This bound depends on the mass ratio
$m_\varphi/m_1$.
The red region is excluded by the AMS positron observation as discussed above.
In addition, we should note that when DM and the mediator $\varphi$ are extremely
degenerate as less than $1\%$, direct searches for DM may affect to the
analysis even if a leptophilic DM is concerned because an interaction
between DM and quarks is derived at one-loop level. 
For Majorana DM, only anapole interaction $a$ which is given by 
\begin{equation}
\mathcal{L}_{\mathrm{eff}}=a\overline{\chi}\gamma^\mu\gamma_5\chi\partial^{\nu}F_{\mu\nu},
\end{equation}
is relevant for direct detection of DM where $F_{\mu\nu}$ is the electromagnetic field
strength.\footnote{When a Majorana DM and an excited DM
$\chi_i~(i\neq1)$ are degenerate as in our framework, transition
dipole interactions among them may be obtained if CP phase exists in the
Yukawa coupling~\cite{Schmidt:2012yg}. However in this paper, since we assumed
real Yukawa couplings, the transition dipole moment does not exist.} 
The concrete expression of the anapole moment and the related
loop function have been obtained
in ref.~\cite{Kopp:2014tsa}. 
The elastic cross section with nuclei derived from the anapole
moment is suppressed by the DM velocity squared. 
However according to ref.~\cite{Kopp:2014tsa}, when DM $\chi_1$ and
the charged mediator $\varphi$ are extremely degenerate as less than a few
$\%$, direct searches for DM would give a constraint on the cross
section for internal bremsstrahlung due to the existence of a pole of the anapole moment at
$m_\varphi/m_1=1$. Enhancement of the anapole moment is especially large when DM
interacts with electron, not but muon and tau since the anapole moment
is proportional to $\log(m_\ell^2/m_\chi^2)$ for $|q|^2\ll m_\ell^2$
where $|q|\lesssim m_\mu$ is the transfer momentum of photon. For
$|q|^2\gg m_\ell^2$, 
this factor is replaced by $\log(|q|^2/m_\chi^2)$. Thus the strongest constraint
is obtained for electron. 
The green region in Fig.~\ref{fig:exp} is excluded by the LUX experiment
where the coupling with electron is assumed. 
In particular, the mass degeneracy between DM and the mediator is
$m_\varphi/m_1=1.01$, a lower DM mass region expected to be tested by
the future gamma-ray experiments is excluded.

DM and the charged scalar masses may be constrained by the LHC as
well since a
pair of the charged scalar $\varphi$ can be produced via the Drell-Yan process. 
The bound for slepton search via the channel
$q\overline{q}\to\gamma/{Z^*}\to\tilde{\ell}\tilde{\ell}^\dag\to\ell\overline{\ell}+E\!\!\!/$
would be applied for our case as good approximation because the
charged scalar $\varphi$ can be regarded as a kind of slepton. 
The ALTAS and CMS Collaborations have analyzed $20.3~\mathrm{fb}^{-1}$
and $19.5~\mathrm{fb}^{-1}$ of the LHC data at $\sqrt{s}=8~\mathrm{TeV}$
respectively, and the constraints for the DM and slepton masses have been
derived~\cite{Aad:2014vma, Khachatryan:2014qwa}. 
However since we are interested in degenerate mass region between DM and
the charged scalar, the constraint of the LHC is relaxed and there is
no substantial constraint for $m_\varphi/m_1\lesssim1.2$. 
Note that if the charged lepton $\ell$ in our framework is replaced to a quark, much
stronger constraint is imposed by the LHC.

%%%%%%%%%%%%%%%%%%%%%%%%%%%%%%%%%%%

%%%%%%%%%%%%%%%%%%%%%%%%%%%%%%%%%%%
\section{Conclusions and Discussions}
\label{sec:4}
Identifying DM is one of the primary issues in (astro-)particle physics. 
Looking for line like gamma-rays coming from the galaxy is important
for DM searches since some DM candidates can generate sharp gamma-rays
which  are not expected to be induced by astrophysical sources. In
particular, internal bremsstrahlung of DM shows an 
interesting sharp gamma-ray spectrum. 
We have considered internal bremsstrahlung for leptophilic Majorana DM with
degenerate particles. When the degenerate particles have the Yukawa
coupling which is smaller than that for DM, we need a larger Yukawa
coupling for DM to satisfy the measured relic density. As a result, the
cross section for internal bremsstrahlung has been enhanced. 
More than one order of magnitude of increase has been achieved for some parameter
sets, and it can be testable by the future gamma-ray experiments. 
We have also considered the constraints from perturbativity of the DM Yukawa
coupling, the AMS positron observation and direct search for DM via anapole moment. 
While the elastic cross section with nucleon derived from anapole
moment is suppressed by the DM velocity, a parameter space has
been excluded because of huge enhancement
of the anapole moment at $m_\varphi/m_1\approx1$.

Finally we briefly comment on the other aspects. 
Small mass difference as we have considered would be induced by
introducing an extra $U(1)$ or a flavor symmetry. 
For example, one can construct a model that one Dirac fermion DM with a
charge of the extra $U(1)$ symmetry is split to two degenerate Majorana fermions after
symmetry breaking. 
Furthermore, the leptophilic DM we have considered in this paper can be
identified as the TeV scale right-handed neutrino. Such a TeV or
electroweak scale right-handed neutrino is included in some models with radiative neutrino
masses. Thus our framework discussed in this paper naturally 
works in these models. 

%%%%%%%%%%%%%%%%%%%%%%%%%%%%%%%%%%%

\section*{Acknowledgments}
T. T. acknowledges support from the European ITN project
(FP7-PEOPLE-2011-ITN, PITN-GA-2011-289442-INVISIBLES) and P2IO
Excellence Laboratory (Labex).

%%%%%%%%%%%%%%%%%%%%%%%%%%%%%%%%%%%
%%%%%%%%%%%%%%%%%%%%%%%%%%%%%%%%%%%

\end{document}